\documentclass[twocolumn,prb,aps,showpacs,floatfix,superscriptaddress]{revtex4}
\usepackage{amsmath,amssymb,graphicx}

\begin{document}

\newcommand{\crb}{CrB$_{2}$}
\newcommand{\rxx}{$\rho_{xx}$}
\newcommand{\rxy}{$\rho_{xy}$}

\newcommand{\hto}{Ho$_{2}$Ti$_{2}$O$_{7}$}
\newcommand{\dto}{Dy$_{2}$Ti$_{2}$O$_{7}$}
\newcommand{\tto}{Tb$_{2}$Ti$_{2}$O$_{7}$}

\newcommand{\nmo}{Nd$_{2}$Mo$_{2}$O$_{7}$}
\newcommand{\pio}{Pr$_{2}$Ir$_{2}$O$_{7}$}

\title{Low-temperature properties of single-crystal CrB$_{2}$}

\author{A.~Bauer}
\affiliation{Physik-Department, Technische Universit\"at M\"unchen, D-85748 Garching, Germany}

\author{A.~Regnat}
\affiliation{Physik-Department, Technische Universit\"at M\"unchen, D-85748 Garching, Germany}

\author{C.G.F~Blum}
\affiliation{Leibniz Institute for Solid State and Materials Research IFW, D-01171 Dresden, Germany}

\author{S.~Gottlieb-Sch\"onmeyer}
\affiliation{Physik-Department, Technische Universit\"at M\"unchen, D-85748 Garching, Germany}

\author{B.~Pedersen}
\affiliation{Heinz Maier-Leibnitz Zentrum (MLZ), D-85748 Garching, Germany}

\author{M.~Meven}
\affiliation{Institut f\"ur Kristallographie, RWTH Aachen, Outstation at MLZ, D-85747 Garching, Germany}
\affiliation{J\"ulich Centre for Neutron Science, Forschungszentrum J\"ulich GmbH, Outstation at MLZ, D-85747 Garching, Germany}

\author{S.~Wurmehl}
\affiliation{Leibniz Institute for Solid State and Materials Research IFW, D-01171 Dresden, Germany}
\affiliation{Institut f\"ur Festk\"orperphysik, Technische Universit\"at Dresden, D-01062 Dresden, Germany}

\author{J.~Kune\v{s}}
\affiliation{Institute of Physics, Academy of Sciences, Praha 6 16253, Czech Republic}

\author{C.~Pfleiderer}
\affiliation{Physik-Department, Technische Universit\"at M\"unchen, D-85748 Garching, Germany}


\begin{abstract}
We report the low-temperature properties of $^{11}$B-enriched single-crystal {\crb} as prepared from high-purity Cr and B powder by a solid-state reaction and optical float zoning. The electrical resistivity, $\rho_{\rm xx}$, Hall effect, $\rho_{\rm xy}$, and specific heat, $C$, are characteristic of an exchange-enhanced Fermi liquid ground state, which develops a slightly anisotropic spin gap $\Delta \approx 220\,{\rm K}$ below $T_{\rm N}=88\,{\rm K}$. This observation is corroborated by the absence of a Curie dependence in the magnetization for $T\to0$ reported in the literature. Comparison of $C$ with $d\rho_{\rm xx}/dT$, where we infer lattice contributions from measurements of VB$_2$, reveals strong antiferromagnetic spin fluctuations with a characteristic spin fluctuation temperature $T_{\rm sf}\approx 257\,{\rm K}$ in the paramagnetic state, followed by a pronounced second-order mean-field transition at $T_{\rm N}$, and unusual excitations around $\approx T_{\rm N}/2$. The pronounced anisotropy of $\rho_{\rm xx}$ above $T_{\rm N}$ is characteristic of an easy-plane anisotropy of the spin fluctuations consistent with the magnetization. The ratio of the Curie-Weiss to the N$\acute{\rm{e}}$el temperatures, $f=-\Theta_{\rm CW}/T_{\rm N}\approx 8.5$, inferred from the magnetization, implies strong geometric frustration. All physical properties are remarkably invariant under applied magnetic fields up to 14\,T, the highest field studied. In contrast to earlier suggestions of local-moment magnetism our study identifies {\crb} as a weak itinerant antiferromagnet par excellence with strong geometric frustration.

\end{abstract}


\vskip2pc

\maketitle

\section{General Motivation}
\label{motivation}

A variety of microscopic mechanisms may lead to low magnetic transition temperatures in materials with strong interactions. In local-moment insulators well-known examples include competing interactions or the effects of geometric frustration. This compares with itinerant-electron magnets, where the coupling of collective spin excitations to the particle-hole continuum may cause low transition temperatures. In fact, one of the oldest problems in condensed matter magnetism may be related the observation of properties strictly believed to be the hallmarks of either local-moment or itinerant-electron magnetism in the same material. Extensive studies of so-called weak itinerant-electron ferromagnets such as ZrZn$_2$, Ni$_3$Al, or MnSi paved the way to a self-consistent quantitative model, which takes into account dispersive spin fluctuations associated with the damping due to the particle-hole continuum. This model reconciles itinerant-electron behavior with features normally attributed to local-moment behavior~\cite{Lonzarich:JPSS1985,Moriya:book}.

Since the early 1990s this self-consistent spin fluctuation model of ferromagnetic $d$-electron systems has been extended to become the basis for studies of quantum phase transitions~\cite{Stewart:RMP2001,HvL:RMP2007}, magnetically mediated superconductivity in selected $f$-electron heavy-fermion systems~\cite{Pfleiderer2009}, and the identification of novel metallic behavior both in intermetallic compounds and transition metal oxides~\cite{Pfleiderer2001,Borzi:Science2007}. Perhaps most important are well defined singularities in the bulk and transport properties of weak itinerant ferromagnets at quantum criticality, which are characteristic of a marginal Fermi liquid breakdown. Yet, these singularities have so far only been observed in the weakly ferromagnetic state~\cite{Smith2008,Niklowitz:2005,Steiner:2003}.

In comparison to $d$-electron ferromagnets the experimental situation in antiferromagnetic $d$-electron systems is much less clear, providing the main motivation for our study. Notably, while systems such as ZrZn$_2$, Ni$_3$Al, and MnSi provide a weakly magnetic counterpart to Fe, Ni, and Co that allows to test theory carefully, equivalent weak itinerant $d$-electron antiferromagnets have been extremely scarce. For instance, the Heusler compound Mn$_3$Si is a good metal that displays a low N$\acute{\rm{e}}$el temperature $T_{\rm N}=23\,{\rm K}$, but the magnetic order is rather complex consisting of a combination of small and large ordered moments~\cite{Tomiyoshi:JPSJ1975,Pfleiderer2003}. Likewise, the antiferromagnetic order below $T_{\rm N}=50\,{\rm K}$ in the metallic semi-Heusler compound CuMnSb displays large ordered moments and a combination of pronounced local-moment and itinerant-electron properties~\cite{Doerr2004,Boeuf2006}. A third candidate material is NiS$_2$, which is, however, an electrical insulator at ambient pressure that exhibits strong geometric frustration~\cite{Niklowitz2008}.

In fact, the metallic state in the presence of geometrically frustrated magnetic interactions touches on a wide range of questions that have recently generated great interest~\cite{Gardner:2010}. For instance, rare-earth pyrochlore insulators such as {\dto} and {\hto} display so-called spin ice behavior at low temperatures, where spin-flip excitations have been interpreted as deconfined magnetic monopoles~\cite{Bramwell:2001,Sakakibara:2003,Castelnovo:2008,Fennel:2009,Morris:2009,Krey:2012}. In systems such as {\tto} strong quantum fluctuations have inspired predictions of enigmatic states such as quantum spin ice \cite{Molavin:2009,Legl:2012}. The interplay of these magnetic excitations with itinerant electrons is essentially unexplored. In metallic pyrochlore systems such as {\nmo} topological contributions to the Hall effect have been identified~\cite{Taguchi:2001}, whereas pyrochlore iridates such as {\pio} are candidate materials for a generic non-Fermi-liquid breakdown~\cite{Machida:2010}. This underscores the potential of metallic antiferromagnets with strong geometric frustration as an important opportunity for scientific exploration.

In turn the identification of metallic $d$-electron systems with weak antiferromagnetic order potentially allows to address questions such as, Are there weakly antiferromagnetic materials consistent with the framework of the self-consistent spin fluctuation theory of ferromagnetic materials? What can be said about the character and importance of spin fluctuations in such weak itinerant antiferromagnets? Do weak itinerant antiferromagnets display singularities in their bulk and transport properties characteristic of marginal Fermi liquid behavior as observed for their weakly ferromagnetic siblings? What are the consequences of strong geometric frustration for the nature of the metallic state of weakly antiferromagnetic materials?

In this paper we address the properties of {\crb}, for which antiferromagnetism with a comparatively low transition temperature $T_{\rm N}\approx88\,{\rm K}$ was reported already in the late 1960s~\cite{Barnes1969,Castaing1969}. To date essentially all studies reported in the literature~\cite{Barnes1969,Castaing1969} have examined the properties of polycrystalline samples with the exception of two short papers on the magnetic and electronic properties of single crystals~\cite{Tanaka1976,Balakrishnan2005}. It is helpful to summarize key questions these studies have either not clarified or not addressed (a detailed account of the literature will be given in Sec.~\ref{introduction}), namely, 
as there are claims of itinerant and local-moment antiferromagnetism, what is the nature of the antiferromagnetic order of {\crb}? What is the effect of magnetic fields on the antiferromagnetic transition temperature in {\crb} and what does the magnetic phase diagram of {\crb} look like (so far no magnetic field studies have been reported at all)? In view of the hexagonal crystal structure of {\crb} how important is geometric frustration? What qualitative as well as quantitative experimental evidence is there for spin fluctuations and their anisotropy in the bulk and transport properties? What is the nature of the metallic state in {\crb}, and does the observation of a strong increase of the susceptibility for $T\to0$ reported in essentially all previous studies provide evidence for marginal Fermi liquid behavior akin to that seen in the ferromagnetic systems? What is the nature of the antiferromagnetic phase transition; i.e., is it first or second order? Is the antiferromagnetic state characterized by a gap, and if so, what is the size, anisotropy, and character of this gap? And last but not least, what is the character and strength of the magnetic anisotropies?

For our study we have prepared large single crystals of {\crb} by optical float zoning, where we exploited a solid-state reaction as an intermediate step in the preparation. Characterizing these samples by the usual set of metallurgical methods including neutron diffraction as well as physical properties such as the resistivity, magnetization, and specific heat, we find by far the highest sample quality reported in the literature to date. Performing high-resolution measurements of the electrical resistivity, Hall effect, magnetization, and specific heat we are able to clarify several of the questions listed above. Our main results may be summarized as follows. 

The electrical resistivity, $\rho_{\rm xx}$, and Hall effect, $\rho_{\rm xy}$, are characteristic of a pure metallic state with large charge carrier mean-free paths approaching $\sim10^3\,{\rm \AA}$ in the zero-temperature limit. Both $\rho_{\rm xx}$ and $\rho_{\rm xy}$ are thereby characteristic of an exchange-enhanced Fermi liquid with a moderately enhanced slightly anisotropic quadratic temperature dependence $\rho_{\rm xx}\propto AT^2$ that develops a slightly anisotropic spin gap $\Delta$ below $T_{\rm N}=88\,{\rm K}$. This evidence for an anisotropic spin-gapped Fermi liquid ground state, which has not been reported before, is corroborated further by the specific heat, which displays to leading order a moderately correlation-enhanced contribution to the Sommerfeld coefficient $\gamma_{\rm corr}\approx9\,{\rm mJ\,mol^{-1}K^{-2}}$ consistent with the Kadowaki-Woods ratio. Further, in our samples we do not observe the ubiquitous divergence of the magnetization for $T\to0$, reported in essentially all previous studies. Instead our study suggests that the divergence in $M$ reported previously must be an impurity effect. This rules out an incipient form of a marginal Fermi liquid ground state akin to that reported for the weak itinerant ferromagnets.

Careful comparison of the temperature dependence of the specific heat, $C$, with the derivative of the electrical resistivity, $d\rho_{\rm xx}/dT$, estimating lattice contributions from measurements of VB$_2$, reveal a pronounced second-order mean-field anomaly at $T_{\rm N}$ not reported before. Based on the high density of our data, which goes well beyond previous work, the specific heat and an analysis of the associated temperature dependence of the entropy reveal strong signatures of an abundance of antiferromagnetic spin fluctuations with a characteristic spin fluctuation temperature $wT_{\rm sf}\approx 257\,{\rm K}$ as well as unusual additional excitations halfway below $T_{\rm N}$. The pronounced anisotropy of the electrical resistivity by up to a factor of two well above $T_{\rm N}$, also reported here, suggests strongly that the spin fluctuations are anisotropic, consistent with the weak easy-plane magnetic anisotropy inferred from the magnetization.

An aspect not considered in the literature concerns the importance of geometric frustration, where our magnetization data reveal a large ratio of the Curie-Weiss to the N$\acute{\rm{e}}$el temperatures, $f=-\Theta_{\rm CW}/T_{\rm N}\approx 8.5$. Thus the effects of geometric frustration are strong. Studying the magnetic field dependence, we find that all physical properties of {\crb} investigated in our study are remarkably invariant up to 14\,T, the highest field studied, as compared with the resolution of our measurements, which is much smaller than this magnetic field scale. In view of the evidence for an abundance of soft spin fluctuations this may be the most remarkable property.

Taken together, the specific temperature dependences and quantitative values of all properties studied, in particular the striking lack of magnetic field dependence, unambiguously identify {\crb} as a material stabilizing weak \textit{itinerant} antiferromagnetism par excellence, clearly contrasting claims in the literature of local-moment magnetism. Hence {\crb} may be considered the perhaps  closest antiferromagnetic analog of the class of weak itinerant ferromagnets so far with the added bonus that {\crb} promises insights into the interplay of geometric frustration and weak itinerant-electron magnetism in a pure metal.


\section{Introduction}
\label{introduction}

{\crb} belongs to the series of hexagonal C32 diborides, $M$B$_2$ (space group $P6/mmm$), where $M$ is a transition-metal or rare-earth element. These systems have first attracted interest due to their high mechanical and thermal stability, high chemical inertness, as well as high electrical and thermal conductivities~\cite{Gordon:JPCS1975,Tian:JPCM1992}. In recent years the C32 diborides have additionally attracted scientific interest because they display a wide range of unusual electronic and magnetic properties. Hence, the C32 diborides offer a unique opportunity to trace the emergence of a wide range of different properties in a crystallographic environment with essentially unchanged unique characteristics. Before turning to the properties of {\crb} it is helpful to review the salient features of the C32 diborides.

\begin{figure}
\includegraphics[width= 0.65\linewidth,clip=]{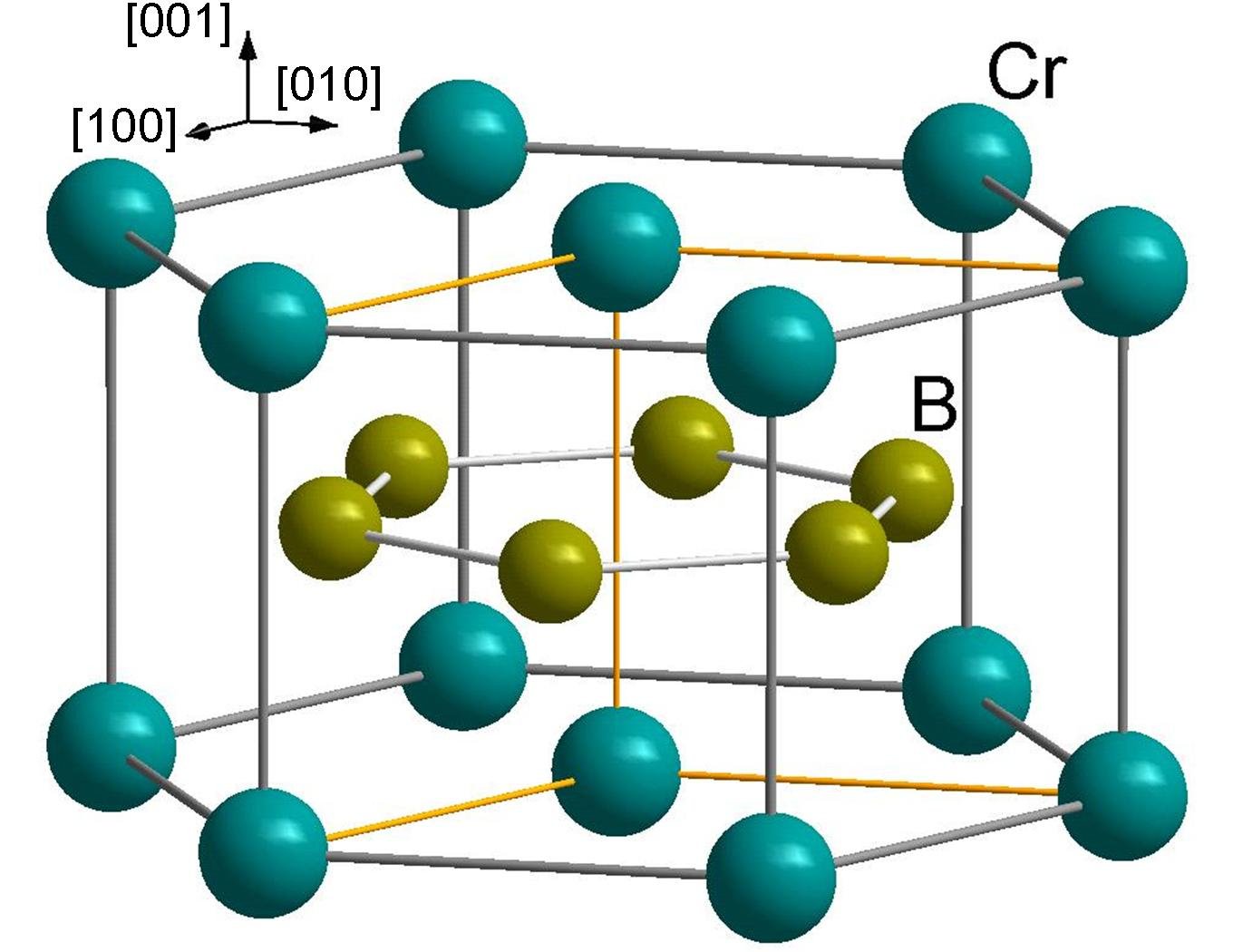}
\caption{(Color online) Schematic picture of the crystal structure of the C32 diborides for the case of {\crb}.}
\label{figure01}
\end{figure}

The crystal structure of the C32 diborides is depicted in Fig.~\ref{figure01}. It is characterized by an alternating sequence of closest-packed $M$ layers and honeycomb B layers along the $\left[001\right]$ direction (the B layers are isostructural to graphite). Other than for instance in the cuprate superconductors the layering displays strong interlayer interactions and represents essentially a dense packing. The B layers form hexagons, where each B has 3 nearest neighbors. In the center above (or below) the B hexagons the transition-metal or rare-earth atoms are located. Each metal ion hence has 12 B nearest neighbors. In turn the structural and electronic properties require consideration of the B-B, $M$-B, and $M$-$M$ bonding. 

Several electronic structure calculations have addressed the chemical bonding in C32 transition metal diborides in view of their rather special structural characteristics~\cite{Liu1975,Armstrong1983,Wang1994,Vajeeston2001}. As a crude approximation and a starting point for thinking about the chemical bonding and band filling it is helpful to distinguish the B $2s$ and $2p$ orbitals in terms of an in-plane $sp^2$ hybrid and an out-of-plane $p_z$ orbital. The strong $\sigma$ bonding of the former results in a large splitting of bonding-antibonding states ($\approx25$\,eV as measured from the bottom to the top of the $sp^2$ manifold). In fact, a gap exists between the $sp^2$ bonding and antibonding bands, in which a narrower $\pi$-bonded B $p_z$ band is situated.

Among the electronic properties of C32 diborides perhaps most widely known is the observation of conventional electron-phonon mediated superconductivity in MgB$_2$~\cite{Nagamatsu2001}. In this compound the Mg $3s$ orbitals donate their 2 electrons to form a broad band above the Fermi level, which is situated close to the top of the $sp^2$ antibonding band. The resulting Fermi surface has a cylindrical shape due to a weak out-of-plane dispersion. The interest in MgB$_2$ is thereby based on two different aspects. First, MgB$_2$ displays a record high $T_{\rm c}=39.5\,{\rm K}$ for an electron-phonon mediated superconductor, which originates in the $E_{2g}$ optical phonon of the in-plane motion of the B atoms that couples very strongly to electrons in the covalent two-dimensional $\sigma$ band. Second, MgB$_2$ represents a showcase for multigap superconductivity in which two almost independent superconducting condensates, one associated with the $\sigma$ and the other with the $\pi$ bands, have different superconducting transition temperatures and distinct superconducting properties~\cite{Canfield2003}. Surprisingly so far, MgB$_2$ appears to be the only superconductor among the C32 diborides (OsB$_{2}$ and RuB$_{2}$ are superconductors but crystallize in the orthorhombic $Pmmn$ structure~\cite{Vandenberg1975,Singh2007,Suh2007}).

In contrast to the hitherto unique example of superconductivity in MgB$_2$, quite a few rare-earth based C32 systems such as TbB$_{2}$, DyB$_{2}$, HoB$_{2}$, ErB$_{2}$, and TmB$_{2}$ order ferromagnetically at low temperatures. Less common is the formation of antiferromagnetic order in rare-earth based systems as observed, for instance, in YbB$_{2}$~\cite{Novikov2010}. This compares with the transition-metal diboride MnB$_{2}$, which displays an antiferromagnetic transition at $T_{\rm{N}} = 670$\,K~\cite{Kasaya1970} followed by a putative ferromagnetic transition at $T_{\rm{C}} = 157$\,K\,\cite{Cadeville1965} (it is possible that the ferromagnetism in MnB$_{2}$ represents, in fact, ferrimagnetism due to a tilting of the antiferromagnetic unit cells). The origin of the magnetic order as well as further electronic instabilities of the electronic structure have so far not been explored in these systems. Namely, in all systems studied to date the question whether the magnetic properties are better described from a local-moment or itinerant-electron point of view is unresolved both from an experimental and a theoretical point of view. Also unexplored in the C32 diborides are the effects of geometric frustration, which may be strong for the hexagonal crystal structure.


We now turn to {\crb}, the topic of this paper, in which early nuclear magnetic resonance (NMR) studies were interpreted as evidence of itinerant antiferromagnetism~\cite{Barnes1969} below $T_{\rm{N}} \approx 88$\,K, representing another important form of electronic order in the class of C32 diborides. However, the formation of itinerant antiferromagnetism was questioned in further NMR studies on powder samples, which suggested that the magnetic order in {\crb} is, in fact, intermediate between archetypal local-moment and itinerant-electron magnetism~\cite{Kitaoka1978,Kitaoka1980}. These ambiguities continued in measurements of the magneto-volume effect, which revealed that the thermal expansion coefficient in the paramagnetic temperature region is nearly the same as in the weak itinerant ferromagnet ZrZn$_{2}$~\cite{Nishihara1987}. Further, the electronic structure of {\crb} has been addressed theoretically in several studies~\cite{Liu1975,Armstrong1983,Wang1994,Vajeeston2001,Fedorchenko2009,Brasse:PRB2013}. Using a non-self-consistent KKR method Liu \textit{et al.} even predicted a magnetic ordering vector parallel to the $c$ axis (the authors refer to their Ref.~[6] reporting putative evidence for $c$-axis order based on neutron scattering)~\cite{Liu1975}.

Neutron scattering provided evidence for cycloidal magnetic order with a small magnetic moment of 0.5\,$\mu_{\rm{B}}\,\rm{f.u.}^{-1}$ at a wave vector $\textbf{q} = 0.285\,\textbf{q}_{110}$, $q_{110} = 2\pi/\frac{a}{2}$~\cite{Funahashi1977}. However, the neutron scattering study had to be performed on a very thin sample to overcome the extremely strong absorption of neutrons due to the large $^{10}$B content permitting detection of a few Bragg reflections only~\cite{neutron-absorption}. Further NMR studies on single crystals recently complemented the information inferred from neutron scattering, suggesting a combination of incommensurate and commensurate spin order in {\crb}~\cite{Michioka2007}. Taken together the question of local-moment or itinerant-electron antiferromagnetism has not been settled.

While a large number of studies addressed the properties of polycrystals there have only been two studies of the electronic and magnetic properties in single-crystal {\crb}~\cite{Tanaka1976,Balakrishnan2005}, one of which addressed the anisotropy of the resistivity and susceptibility~\cite{Tanaka1976}. Moreover, a strong divergence of the magnetization for $T\to0$ has been observed in all previous studies including one of the single-crystal studies~\cite{Balakrishnan2005}, whereas data in the other single-crystal study~\cite{Tanaka1976} are rather sketchy. While this divergence may be a so-called Curie tail due to impurities, there exists also the interesting possibility that the divergence is intrinsic and an indication for an incipient breakdown of the description of the metallic state in terms of a Fermi liquid as observed in the ordered state of weak itinerant ferromagnets. 

To clarify the questions in {\crb} summarized in Sec.~\ref{motivation} we have grown high-purity single crystals. As a congruently melting compound the preparation of single crystals of {\crb} is, in principle, straight-forward~\cite{Portnoi1969}. Accordingly, various techniques have been reported in the literature. In early studies arc melting of appropriate amounts of pure Cr and B was used to prepare polycrystals~\cite{Cadeville1965,Castaing1972,Guy1976}. Strongly textured polycrystals with nearly single-crystal character as well as single crystals were prepared in a hot graphite crucible~\cite{Boeuf2003}. Best results were reported for optically or radio-frequency heated float zoning of polycrystals prepared in various ways~\cite{Tanaka1976,Okada1996,Balakrishnan2005,Michioka2007,Fedorchenko2009}. Yet, two major constraints have limited the sample quality in previous studies. First, due to the high melting temperature of about $2150\,^{\circ}{\rm C}$ the high vapor pressure of B leads to considerable losses. Second, essentially all previous studies displayed fairly pronounced Curie tails below $T_{\rm N}$ that are most likely due to Fe impurities in the starting elements~\cite{Boeuf2003,Balakrishnan2005,Fedorchenko2009}.

For the work reported here and additional studies to be reported elsewhere we have grown large single crystals of {\crb} by a solid-state reaction followed by optical float zoning. We observe consistently in all properties evidence for an excellent sample quality, which we attribute to the use of high-purity starting elements and high Ar pressures of 15\,bars to reduce the loss of B during float zoning. In addition, our samples were grown using 99\,\% enriched $^{11}$B powder to permit detailed neutron scattering studies in the future. Besides the single-crystal preparation and characterization, we report in this paper comprehensive measurements of the electrical resistivity, the Hall effect, the magnetization, and the specific heat at low temperatures under applied magnetic fields. In comparison to previous work our data were recorded at much reduced noise and much higher density, addressing a much wider range of temperatures and the effects of magnetic fields.

Further studies which are presently under way on the same single crystals may be summarized as follows. First, elastic neutron scattering studies at the single-crystal diffractometers RESI and HEiDi at the Heinz Maier-Leibnitz Zentrum in Munich (MLZ) explore the precise nature of the incommensurate spin order addressed first by Funahashi \textit{et al.} in Ref.~\onlinecite{Funahashi1977}. These data await full refinement, where the hexagonal crystal structure leads to considerable undersampling and additional complexities~\cite{Regnat:RESI2013}. Second, inelastic neutron scattering studies at the cold triple-axis spectrometer PANDA at MLZ reveal the presence of an abundance of strongly damped spin excitations both above and below $T_{\rm N}$~\cite{Brandl:neutrons}. However, these studies still await a complex polarization analysis. Third, at temperatures down to 0.3\,K and magnetic fields up to 14\,T the angular dependence of three distinct de Haas--van Alphen (dHvA) frequencies of the torque magnetization have been recorded~\cite{Brasse:PRB2013}. Comparison with the Fermi surface of antiferromagnetic {\crb} calculated in the local density approximation (LDA) using the generalized gradient approximation (GGA) of the exchange and correlation functionals suggests that two of the observed dHvA oscillations arise from electron-like Fermi surface sheets formed by bands with strong B $p_{x,y}$ character, while the third may be related to Cr $d$ bands. Yet, the majority of the heavy $d$ bands could not be detected at the high temperatures and low fields studied.

The calculated electronic structure suggests further, that general features similar to those mentioned above for MgB$_2$ can be found in {\crb}, where the details in the vicinity of the Fermi level are quite different due to the presence of narrow ($\approx4$\,eV) partially filled Cr 3$d$ bands~\cite{Brasse:PRB2013}. The B $p_z$ states are thereby pushed away from the Fermi level while the $sp^2$ bands acquire three-dimensional character leading to a quasispherical Fermi surface. The large density of Cr $d$ states in the vicinity of the Fermi level is, finally, a prerequisite for the magnetic order at low temperatures. These calculations identify {\crb} as a covalent compound reminiscent of Cr metal rather than Cr oxides, i.e., a metal in which a broad 4$s$ band overlaps with narrower 3$d$ bands such that they cannot be distinguished very accurately.

The outline of our paper is as follows. In Sec.~\ref{methods} we present a short summary of the experimental methods, emphasizing key aspects of the single-crystal preparation. This is followed by an account of the results of measurements of the resistivity, the magnetization, and the specific heat in Sec.~\ref{results}. As part of Sec.~\ref{results} data are analyzed and discussed in a basic manner, pointing out evidence for the formation of weak itinerant antiferromagnetism in the presence of strong geometric frustration. The paper concludes with a more general discussion in Sec.~\ref{further}, addressing among other issues the consistency with other correlated materials and Fermi liquid behavior as well as the character and strength of the spin fluctuations.


\section{Experimental methods}
\label{methods}

\subsection{Single Crystal Growth}

For the preparation of our {\crb} single crystals 4N5 chromium powder ($-100+325\,{\rm mesh}$) and 4N5 boron powder ($<140\,{\rm mesh}$) was used. This purity minimizes contamination by Fe impurities reported in the literature as described above, which may be traced back to the Cr and B used in commercially available {\crb} powder. The B powder used was 99\,\% $^{11}$B-enriched to permit comprehensive neutron scattering studies of the magnetic and structural properties~\cite{neutron-absorption} to be reported elsewhere~\cite{Regnat:RESI2013}. As the first step of the preparation, stoichiometric amounts of the Cr and B powder were thoroughly mixed and filled into a bespoke two-component tungsten crucible with an inner diameter of 6\,mm and a length of 90\,mm. The crucible was then mounted on a horizontal water-cooled Cu hearth in an ultrahigh vacuum (UHV) system. After carefully pumping the UHV system the Cr-B powder was heated by radio-frequency induction under an inert atmosphere at a gas pressure of 1.1\,bars, where the 6N Ar had been additionally purified with a hot getter furnace (NuPure Corporation). As the powder reached a temperature of $\sim1500\,^{\circ}{\rm C}$ a metallic rod formed through an exothermal solid-state reaction. Heating and cooling of the tungsten crucible for this solid-state reaction was carried out in less than ten minutes. The sintered {\crb} rod had an estimated density of $\sim2.6\,{\rm g\,cm}^{-3}$, i.e., $\sim 50\%$ of the density of crystalline {\crb}. This proved to be sufficient for the subsequent float zoning.

\begin{figure}
\includegraphics[width= 1.0\linewidth,clip=]{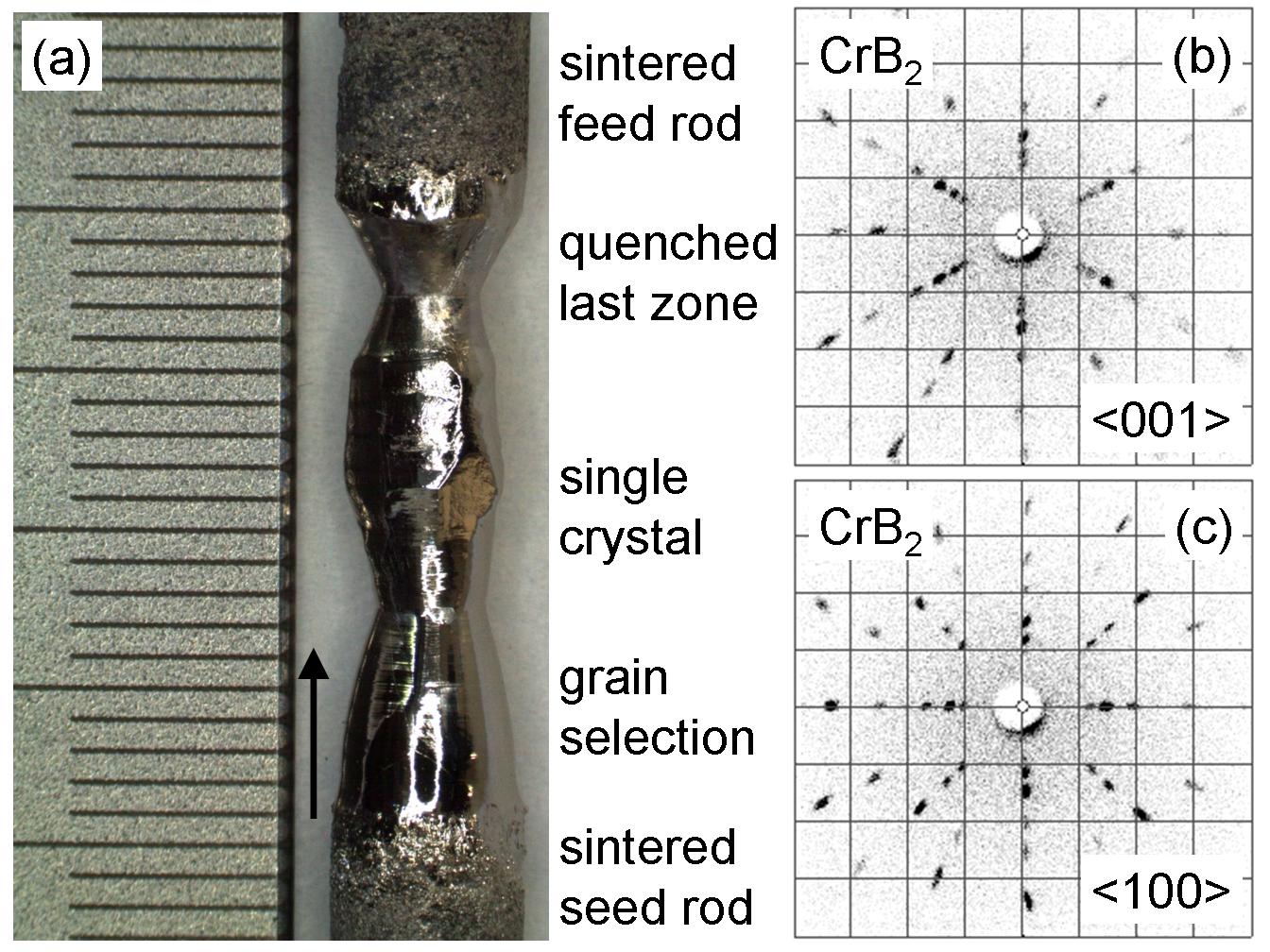}
\caption{(Color online) (a)~{\crb} rod after optical float zoning. Large facets and a clean metallic surface indicate high sample quality. The growth direction was from the bottom to the top. (b)~Laue x-ray diffraction pattern along the $c$ axis showing the characteristic sixfold symmetry. (c)~Laue x-ray diffraction pattern along the $a$ axis of the same part of the ingot. The characteristic twofold symmetry demonstrates that the single-crystal grain extends across the ingot.}
\label{figure02}
\end{figure}

Single-crystal growth was carried out by means of optical float-zoning of the sintered {\crb} rods. Following initial tests with an UHV-compatible four-mirror image furnace\cite{Neubauer2011} the float zoning was eventually performed with a high-pressure crystal growth furnace---Smart Floating Zone (SFZ) at IFW Dresden~\cite{ScIDre}. In the SFZ image furnace light of a 7\,kW xenon arc lamp is focused on the sample by two ellipsoidal mirrors. After evacuating the SFZ system the growth process was carried out under a flowing Ar atmosphere of 15\,bars. The 5N argon was additionally purified by a Ti getter furnace, yielding a measured oxygen content below 0.1\,ppm. Seed and feed rod were counterrotating at rates of 8.5\,rpm and 8.5 to 13\,rpm, respectively. The temperature of the molten zone was measured with a two-color pyrometer using a stroboscopic method~\cite{Behr2007}. The temperature at the surface of the zone was $\sim 2200\,^{\circ}{\rm C}$ during growth in agreement with the binary phase diagram reported in the literature~\cite{Okamoto2003}.

Figure~\ref{figure02}(a) shows the float-zoned {\crb} crystal, which exhibits a shiny metallic surface and well-developed large facets in the regime of the single crystal. After the first 5\,mm of growth complete grain selection had taken place resulting in a single crystal grain across the entire rod. Shown in Figs.~\ref{figure02}(b) and \ref{figure02}(c) are typical Laue x-ray pictures of surfaces perpendicular to the $c$ axis and the $a$ axis, respectively. While the picture along the $c$ axis displays the characteristic sixfold symmetry, the $a$ axis appears to show a fourfold symmetry. Closer inspection confirms, however, the expected twofold symmetry as the lattice constants $a$ and $c$ in {\crb} are very similar. Neutron single-crystal diffraction at the diffractometer RESI and HEiDi at FRM II revealed good single crystallinity of our {\crb} sample. The hexagonal lattice constants of {\crb} observed in neutron scattering are $a = 2.972\,\rm{\AA}$ and $c = 3.083\,\rm{\AA}$ in close agreement with values reported in the literature~\cite{Vajeeston2001}.

Two {\crb} crystals were grown: SFZ118 denotes the single crystal depicted in Fig.~\ref{figure02}(a). Samples from this ingot were used for the bulk measurements presented in Secs.~\ref{results}A, \ref{results}C, and \ref{results}D. A second growth, SFZ162, yielded several large grains that, however, exhibit small-angle grain boundaries. Samples from SFZ118 and SFZ162 were used for the resistivity and Hall effect measurements which are essentially identical (cf. Table~\ref{resistivity-parameter}), where data of SFZ162 are shown in Sec.~\ref{results} B. Moreover, we prepared a sample of VB$_2$ using 99\,\% enriched $^{11}$B to obtain a nonmagnetic, metallic reference compound for our specific-heat measurements, where we used the same crystal growth procedure as for {\crb}. However, due to the very high melting temperature of $\sim2750^{\circ}{\rm C}$ the sintered seed and feed rods of VB$_2$ could not be properly float zoned. Nevertheless we obtained a strongly textured, nearly single-crystalline sample of VB$_2$. Laue x-ray diffraction established several grains in this piece with a small orientational mismatch of a few degrees.

\subsection{Bulk Properties}

For measurements of the electrical transport and thermodynamic bulk properties the {\crb} samples were cut using a wire saw and carefully polished. All bulk measurements (magnetization, ac susceptibility, and specific heat) were performed on a single-crystalline cuboid of $2.5\times2.2\times0.9\,\rm{mm}^{3}$ with $\langle001\rangle\times\langle100\rangle\times\langle210\rangle$ orientations. For the electrical transport and the Hall effect measurements thin platelets of $1\times0.5\times0.3\,\rm{mm}^{3}$ and of $2.0\times1.0\times0.2\,\rm{mm}^{3}$ were prepared, respectively. The platelets had orientations of $\langle001\rangle\times\langle210\rangle\times\langle100\rangle$ for current along the hexagonal $c$ axis and of $\langle100\rangle\times\langle210\rangle\times\langle001\rangle$ for current along the hexagonal $a$ axis, respectively. Gold wires of 25\,$\mu$m diameter were spot welded onto the samples for applying the excitation current and for the voltage pick up. The geometry factors were determined from digital photographs recorded with an optical microscope. A conservative estimate of resulting uncertainty is $\sim$25\%.

The electrical resistivity, {\rxx}, was measured in a cryogen-free adiabatic demagnetization cryostat at temperatures down to $\sim 100\,{\rm mK}$. A digital lock-in technique was used at an excitation frequency of 22.08\,Hz and at low excitation currents to avoid parasitic signal pickup using a four-terminal set-up. A separate set of measurements was conducted in a 14\,T superconducting magnet system at temperatures down to $\sim2.3\,{\rm K}$ using a six-terminal configuration for simultaneous measurements of the longitudinal resistivity and the Hall effect. The Hall resistivity, {\rxy}, was inferred from the transverse voltage pickup, and corresponds to the antisymmetric signal contribution under magnetic field. For details on the method of antisymmetrizing the transverse voltage pickup we refer to the supplement of Ref.~\onlinecite{Ritz:Nature2013}. The longitudinal resistivity obtained in the four- and six-terminal measurements were in excellent agreement.

The magnetization was measured in an Oxford Instruments vibrating sample magnetometer with an oscillation frequency of 62.35\,Hz and an amplitude of roughly 1\,mm. The specific heat was finally measured in a Quantum Design physical properties measurement system with a standard heat-pulse method. Heat pulses, if not stated otherwise, had an amplitude of 0.5\% of the current temperature. Both cryostats provided temperatures down to $\sim2$\,K and magnetic fields up to 9\,T. 


\section{Experimental results}
\label{results}

\begin{figure}
\includegraphics[width= 0.8\linewidth,clip=]{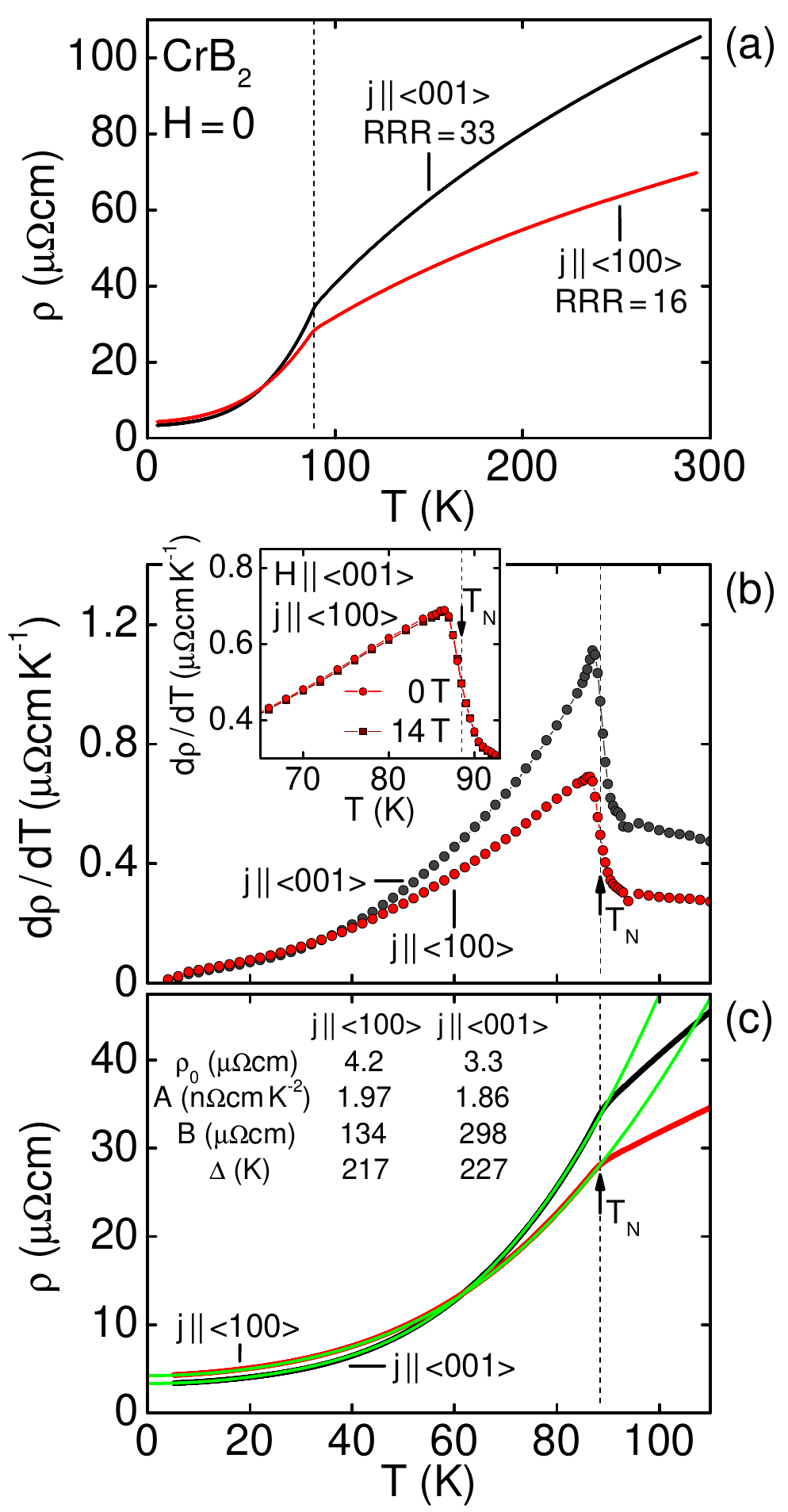}
\caption{(Color online) Temperature dependence of the electrical resistivity of {\crb}. (a)~Electrical resistivity for current parallel to the $a$ axis and the $c$ axis, where a pronounced kink marks the antiferromagnetic transition. (b)~Derivative of the resistivity with respect to the temperature. (c)~Close-up view of the resistivity below $T_{\rm N}$ for both current directions as fitted with a model according to Eq.~\eqref{rho} combining Fermi liquid behavior and scattering by spin waves with a slightly anisotropic spin wave gap.}
\label{figure03}
\end{figure}

\subsection{Resistivity}
\label{resistivity}

Shown in Fig.~\ref{figure03} are typical resistivity data of our {\crb} single crystal SFZ162 in zero magnetic field (data for SFZ118 are essentially identical). With decreasing temperature the resistivity decreases monotonically for both current directions limiting to small residual values, $\rho^{a}_0=4.2\,{\rm \mu\Omega cm}$ and $\rho^{c}_0=3.3\,{\rm \mu\Omega cm}$, for current parallel to the $a$-axis and the $c$-axis, respectively. Our samples exhibit residual resistivity ratios (RRR) of 16 and 32 for current along the $a$ and $c$ directions, respectively. To the best of our knowledge, these values of RRR are the highest reported in the literature so far. With increasing temperature the resistivity develops a pronounced anisotropy, where the resistivity at room temperature for $j\parallel {\langle001\rangle}$ is a factor of two larger than for $j\parallel {\langle100\rangle}$. Since this observation differs distinctly from the only other data on the resistivity for the $a-$ and $c-$axis reported in Ref.~\onlinecite{Tanaka1976} we have carefully double-checked our measurements and confirmed this observation in several samples taken from the different ingots. 

Considering the temperature dependence in more detail we find that $\rho(T)$ for both axes changes with decreasing temperature abruptly from a sublinear to a super-linear dependence at a pronounced kink that marks the onset of magnetic order. We determine the N$\acute{\rm{e}}$el temperature from the first derivative of the resistivity, $\text{d}\rho / \text{d}T$, as depicted in Fig.~\ref{figure03}(b), where $T_{\rm{N}} = 88.5$\,K is the same for both current directions. This value of $T_{\rm{N}}$ is consistent with the sketchy data reported in previous studies\cite{Barnes1969,Tanaka1976}, where the high density and low noise of our data permit the calculation of derivatives and a detailed analysis not possible before (see Sec.~\ref{magres} for a comment on the effect of magnetic field). The narrow temperature range of the increase in $\text{d}\rho / \text{d}T$ provides additional evidence of the high compositional and structural homogeneity of our samples.

It is interesting to compare the detailed temperature dependence of $\rho(T)$ near $T_{\rm N}$ with the spin-density wave transition in Cr and the onset of the hidden order in the heavy-fermion system URu$_{2}$Si$_{2}$~\cite{McElfresh1987,Dawson1989}. Both systems display a small, well-developed maximum at temperatures preceding the pronounced drop. This has been interpreted as the formation of a superzone gap. In comparison, we do not observe such a maximum in {\crb}.

For $T<T_{\rm{N}}$ the superlinear dependence can be accounted for extremely well by the formation of spin density wave type order as shown in Fig.~\ref{figure03}(c). Here, we consider three contributions: (i)~a constant term $\rho_{0}$ accounting for impurity scattering, (ii)~the quadratic temperature dependence $AT^{2}$ of a Fermi liquid accounting for electron-electron scattering including umklapp processes, and (iii)~an exponential term accounting for the scattering of electrons from bosonic excitations, e.g., spin waves. The latter is obtained by inserting a dispersion $\omega(\textbf{k})$ with a spin wave gap $\Delta$ into the linearized Boltzmann equation for $T \ll \Delta$~\cite{Anderson1980}. As the three terms are due to different scattering mechanisms the scattering rates may be added in the spirit of Matthiessen's rule, giving
\begin{equation} \label{rho}
\rho(T) = \rho_{0} + AT^{2} + B\frac{T}{\Delta}(1 + 2\frac{T}{\Delta})\exp(-\frac{\Delta}{T})
\end{equation}
Fitting our experimental data we find for the parameters $A^{a} = 1.97\,\rm{n}\Omega$cm\,K$^{-2}$ and $A^{c} = 1.86\,\rm{n}\Omega$cm\,K$^{-2}$ for current along the $a$-axis and $c$-axis, respectively. These values are not uncommon for $d$-band systems as discussed below. They are slightly smaller than the values of $A$ reported by Guy~\cite{Guy1976} for polycrystalline samples, who fitted a quadratic temperature dependence only. The values of the excitation gaps $\Delta^{a} = 217$\,K and $\Delta^{c} = 227$\,K are slightly anisotropic, consistent with the weak magnetic anisotropy seen in the magnetization presented below. 

Fitting the transport data of samples from the second crystal, SFZ118, yields essentially the same set of parameters at low temperatures as summarized in Table~\ref{resistivity-parameter}. While the most likely scenario in {\crb} is that of gapped spin wave excitations, we note that the same expression also accounts for the resistivity of URu$_2$Si$_2$ in the hidden order phase, i.e., a state in which the precise nature of the gapped excitations has been a long-standing mystery.

The sublinear dependence of $\rho(T)$ for $T>T_{\rm N}$ arises from a combination of scattering of the conduction electrons by phonons and an abundance of spin fluctuations as observed in the specific heat presented below. The anisotropy of $\rho(T)$ observed at high temperatures compares thereby with the gap at low temperatures. In turn the resistivity provides evidence that these spin fluctuations are also moderately anisotropic, where the anisotropy (softer fluctuation causing a higher resistivity) is also consistent with the easy-plane anisotropy in the magnetization.\\

\begin{table}
\centering
\caption{
Key parameters inferred from the temperature dependence of the electrical resistivity, when fitting a Fermi liquid ground state that develops a spin wave gap below $T_{\rm N}$. 
\label{resistivity-parameter}
}
\begin{tabular}{cccccc}
\hline\noalign{\smallskip}
\hline\noalign{\smallskip}
&&\multicolumn{2}{c}{SFZ118}&\multicolumn{2}{c}{SFZ162} \\
quantity & unit & \hspace{2mm} $j \parallel a$ \hspace{2mm} & 
\hspace{2mm} $j \parallel c$ \hspace{2mm} & 
\hspace{2mm} $j \parallel a$ & $j \parallel c$ \hspace{2mm} \\
\hline\noalign{\smallskip}
$\rho_0$ & ${\rm \mu\Omega\,cm}$     & 6.5     & 4.1 & 4.2     & 3.3\\
$A$ & ${\rm n\Omega\,cm\,K^{-2}}$     & 1.74     & 2.12     & 1.97     & 1.86\\
$B$ & ${\rm \mu\Omega\,cm}$         & 127     & 355 & 134     & 298 \\
$\Delta$ & K                 & 210     & 226     & 217     & 227\\
RRR & -                 & 11     & 31     & 16 & 32\\
\noalign{\smallskip}\hline
\noalign{\smallskip}\hline
\end{tabular}\\
\end{table}


\subsection{Magnetoresistance}
\label{magres}

In view of the strong spin fluctuations inferred from the resistivity, as well as the specific heat and magnetization presented below, one might expect a strong response to applied magnetic fields rapidly quenching these spin fluctuations. This effect is well established in a large number of $d$- and $f$-electron compounds~\cite{Stewart1984}. Moreover, connected with such a quenching of spin fluctuations one might expect a strong change of the ordering temperature under applied magnetic fields. In stark contrast, we find essentially no change in the temperature dependence of $\rho_{\rm xx}$, in particular near $T_{\rm N}$ where the fluctuations are softest. The lack of field dependence is thereby nicely illustrated in terms of the derivative of $d\rho_{\rm xx}/ dT$ under an applied magnetic field of $14\,{\rm T}$, shown in the inset of Fig.~\ref{figure03}(b). In particular, at 14\,T the change of $T_{\rm N}$ is small than the accuracy of our experiment of a tenth of a percent, i.e., field scale exceeds the detection limit for possible change considerably.

More detailed data of the magnetoresistance up to 14\,T, the highest field studied, reveal conventional, albeit weak, changes of $\rho_{\rm xx}$ as shown in Fig.~\ref{figure04}(a). The magnetoresistance displays thereby a quadratic field dependence $\rho_{\rm xx}\propto B^2$ over nearly the entire field range with a weakly temperature dependent prefactor. In view of the rather complex band structure of {\crb} it is unfortunately not possible to identify the origin of this field dependence at this stage.

\subsection{Hall-Effect}
\label{hall}

\begin{figure}
\includegraphics[width= 0.8\linewidth,clip=]{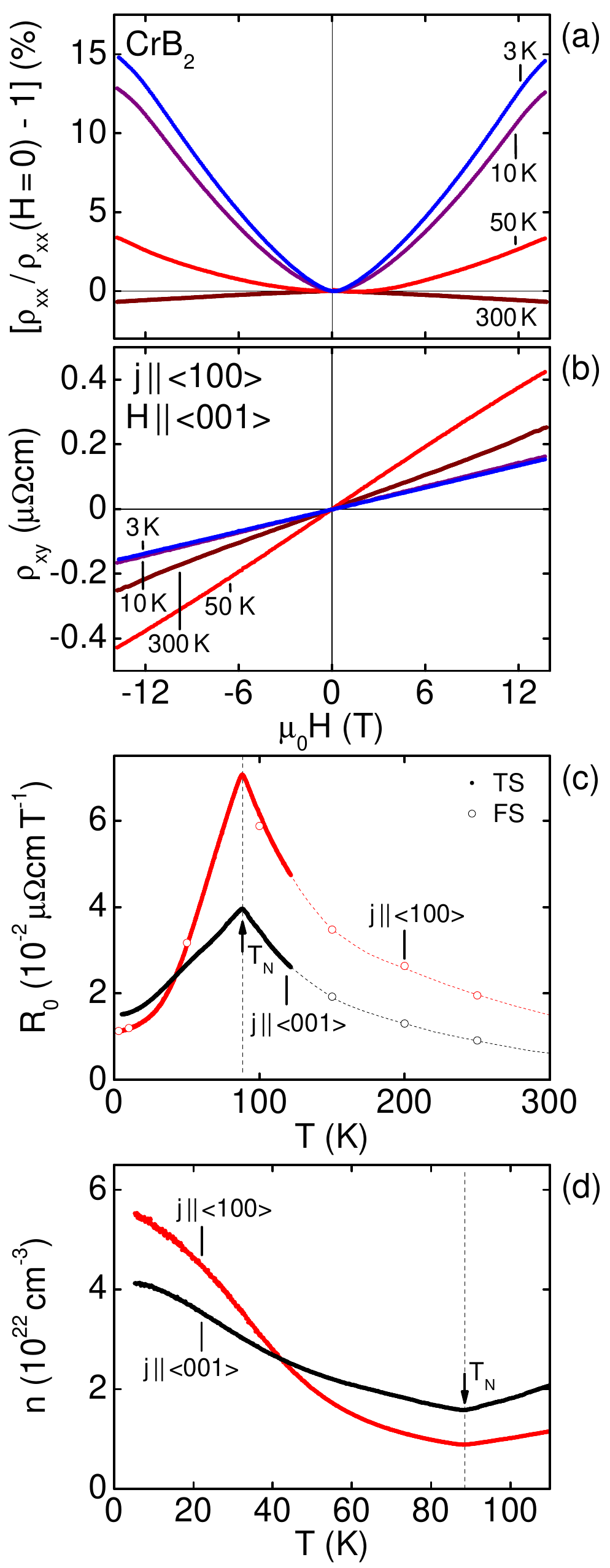}
\caption{(Color online) Magnetoresistance and Hall effect of {\crb}. (a)~Magnetoresistance of CrB$_2$ for various temperatures. (b)~Hall resistivity for various temperatures. (c)~The normal Hall coefficient as a function of temperature for current parallel to the $a$ axis and the $c$ axis. For both directions a pronounced maximum is observed at the N$\acute{\rm{e}}$el temperature $T_{\rm{N}} = 88.5$\,K. Data extracted from measurements as a function of temperature at fixed fields of $\pm6$\,T (solid symbols) and from measurements as a function of field at fixed temperatures (open symbols) are in good agreement. (d)~Effective carrier density as calculated from the data in panel (c).}
\label{figure04}
\end{figure}

The Hall resistivity as measured at selected temperatures displays a linear field dependence up to 14\,T. Essentially no anomalous Hall contributions are found. Typical data recorded at selected temperatures are shown in Fig.~\ref{figure04}(b). For detailed information on the temperature dependence we have additionally performed temperature sweeps in $\pm 6\,{\rm T}$ up to $\sim 125\,{\rm K}$. The associated linear Hall coefficient, $R_0$, is shown in Fig.~\ref{figure04}(c), where values of $R_0$ inferred from field sweeps at higher temperatures are shown as individual data points above $\sim 125\,{\rm K}$. The absolute value of $R_0$ is consistent with previous studies, which reported the Hall coefficient for one fixed field value of $4\,{\rm T}$ without any  information on the field dependence~\cite{Tanaka1976}. We note that our data have been recorded at much lower noise and much higher density as compared to previous work, providing detailed information as a function of temperature and field.

For increasing temperature the Hall coefficient $R_{0}$ increases and displays a sharp cusp at the N$\acute{\rm{e}}$el temperature $T_{\rm{N}} = 88.5$\,K, followed by a decrease. The Hall coefficients for field in the easy plane ($j\parallel \langle 001\rangle$) and along the hard axis ($j\parallel \langle 100\rangle$) are similar. The temperature dependencies cross around 30\,K such that the former is about a factor of two larger at and above $T_{\rm N}$. It is also instructive to calculate the charge carrier concentration $n$ from the temperature dependence of $R_{0}$ as shown in Fig.~\ref{figure04}(d). The positive value of $R_{0}$ is characteristic of electron conduction with an absolute value characteristic of a good metallic state. The increase of $n$ with decreasing temperature below $T_{\rm N}$ suggests thereby that multiple bands, presumably with both electron-like and hole-like character, are present at the Fermi surface and affected differently as a function of temperature~\cite{Brasse:PRB2013}.

We finally note that the temperature dependence shows a faint S shape with a point of inflection around $\sim40\,{\rm K}$ that may correspond to an additional temperature dependence in the specific heat presented below in Sec.~\ref{specificheat}, which indicates excitations beyond the simple formation a spin gap. 


\subsection{Magnetization}
\label{magnetization}

\begin{figure}
\includegraphics[width= 0.8\linewidth,clip=]{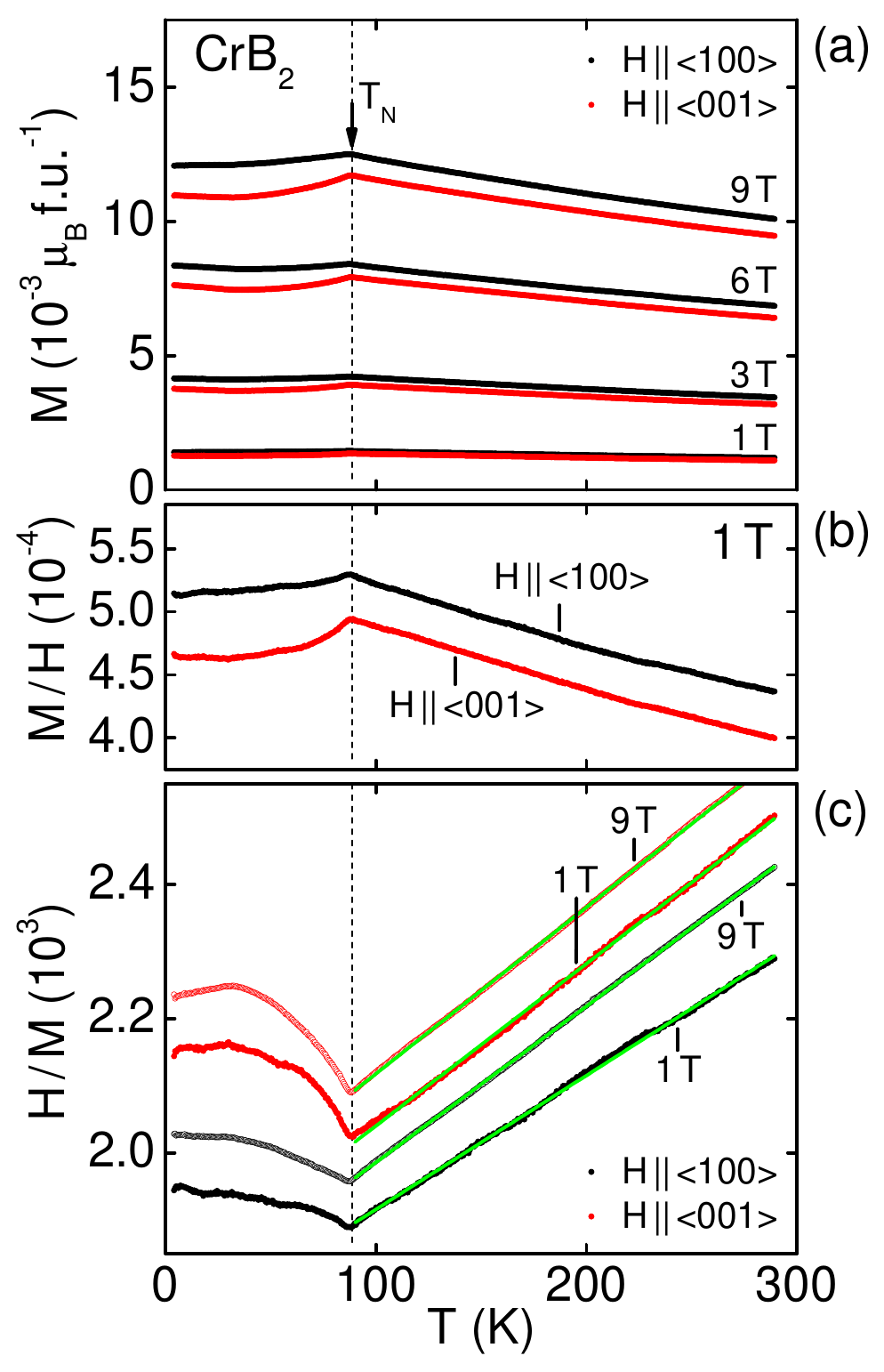}
\caption{(Color online) Magnetization of {\crb} as a function of temperature. (a)~Magnetization for fields up to 9\,T displaying a clear kink at the N$\acute{\rm{e}}$el transition. The absolute values are small. The crystalline $c$ direction corresponds to the magnetic hard axis. No qualitative changes are observed up to the highest fields studied. (b)~Normalized magnetization, $M/H$, for $\mu_{0}H = 1$\,T for both field directions. At low temperatures there is no sign of a Curie tail. (c)~Inverse normalized magnetization, $H/M$, for both field directions as well as for $\mu_{0}H = 1$\,T and $\mu_{0}H = 9$\,T. The solid green lines are Curie-Weiss fits for temperatures $T > 90$\,K.}
\label{figure05}
\end{figure}

The magnetization of our high-quality single crystals, shown in Fig.~\ref{figure05}, allows us to address several important questions. Namely, essentially all previous studies\cite{Boeuf2003,Balakrishnan2005,Fedorchenko2009} have reported a pronounced divergence for $T\to0$ (the only exception is Ref.~\onlinecite{Tanaka1976}). If this divergence were intrinsic, it would perhaps indicate the vicinity to a quantum phase transition and marginal Fermi liquid behavior. For instance, the remnants of such marginal Fermi liquid behavior are indeed observed in the resistivity of the \textit{ordered} state of the class of weak ferromagnets~\cite{Smith2008,Niklowitz:2005,Steiner:2003}. In contrast, in our samples, which by all accounts appear to be of the highest purity achieved so far, we do not observe the ubiquitous upturn of $M$ for $T\to0$ seen before. In fact, it had been suspected that the increase in $M$ represents a Curie tail due to magnetic impurities. For the case of {\crb} the impurities seen in previous studies most likely originated in the Fe content in the Cr and B used for crystal growth.

Itinerant antiferomagnetism in {\crb} is corroborated by the uniform (zero wave vector) magnetization as a function of temperature as measured up to 9\,T. In the paramagnetic state the absolute value of the magnetization is small. Moreover, up to 300\,K the magnetization only varies slightly at a constant field, with a clear cusp at $88.5$\,K. This temperature corresponds very well with the transition temperature observed in the electrical resistivity and specific heat (the latter will be presented below). Both the qualitative shape of the magnetization curve and the position of the maximum are thereby unchanged in fields up to 9\,T for field along the $a$-axis and the $c$-axis. Hence, neither the field dependence (which is not shown) nor the temperature dependence provide evidence of a spin-flop transition. A similarly high stability under very large magnetic fields has been reported for the antiferromagnetic order in Mn$_3$Si and CuMnSb~\cite{Pfleiderer2003,Doerr2004,Boeuf2006}. The large dominant energy scales and very weak coupling between a uniform magnetic field and the finite wave vector (antiferromagnetic) order are an important characteristic of itinerant antiferromagnetism.

Figure~\ref{figure05}(b) shows the normalized magnetization, $M/H$, as a function of temperature, which provides an approximate measure of the susceptibility, $\text{d}M / \text{d}H$. The absolute value is thereby small, positive, and of the order of $5\cdot10^{-4}$ for the parameter range studied. In turn, the magnetization increases linearly for increasing fields. The magnetization and the susceptibility for field along the crystalline $a$ direction are about 10\,\% larger than for field along the $c$ axis. This indicates a very weak easy-plane magnetic anisotropy that corresponds to the hexagonal basal plane of the crystal structure consistent with the rather limited data reported previously~\cite{Tanaka1976}. A recent study by means of measurements of the susceptibility and electronic structure calculations addressed the magnetic anisotropies in several C32 diborides~\cite{Fedorchenko2009}. While the absolute value of the susceptibility of several non-magnetic diborides was reproduced from the first-principles calculations, the magnetic anisotropy and the temperature dependence of the susceptibility in {\crb} were not addressed.

The inverse normalized magnetization, $H/M$, is finally depicted in Fig.~\ref{figure05}(c). It follows a Curie-Weiss dependence between $T_{\rm N}$ and room temperature (the highest temperature measured) for both field directions. The slope of the linear temperature dependence is essentially unchanged in the field range studied. We thereby extrapolate a very large negative Curie-Weiss temperature around $\Theta_{\rm{CW}} = -(750\pm50)$\,K and a large effective fluctuating moment around $\mu_{\rm{eff}} = (2.0\pm0.1)\,\mu_{\rm{B}}\,\rm{f.u.}^{-1}$. The anisotropy of $\Theta_{\rm{CW}}$ is small as compared with uncertainties in extrapolating the value of $\Theta_{\rm{CW}}$. 

The large value of $\Theta_{\rm{CW}}$ implies very large antiferromagnetic interactions. It is hence consistent with the lack of any noticeable field dependence. In fact, we do not expect much field dependence even up to the highest fields currently accessible experimentally of nearly 100\,T. At the same time $\Theta_{\rm{CW}}$ exceeds by a large margin the value of $T_{\rm N}$. The ratio $f=-\Theta_{\rm{CW}}/T_{\rm N}$ is widely considered as a measure of the strength of geometric frustration, where large values of $f$ imply a strong suppression of long-range order and hence strong geometric frustration. In fact, the value of $f\approx 8.5$ we observe in {\crb} is characteristic of strong geometric frustration~\cite{Ramirez1994}.

Finally, the large fluctuating moment exceeds the ordered moment $\mu_{\rm s}\approx 0.5\,\mu_{\rm B}$ inferred in previous NMR studies by a factor of four. A similarly large enhancement is one of the hallmarks of weak itinerant-electron ferromagnets such as MnSi, Ni$_3$Al, or ZrZn$_2$. Neutron scattering studies in these systems in the 1980s eventually revealed the strong damping of collective spin excitations through the coupling to the particle-hole continuum as a new mechanism leading to Curie-Weiss behavior. As an antiferromagnetic system, {\crb} appears to display the largest enhancement of $\mu_{\rm{eff}}/\mu_{\rm s}$ reported to date. This in its own right motivates detailed neutron scattering studies of the spin excitations in {\crb} to be reported elsewhere~\cite{Brandl:neutrons}.

\begin{figure}
\includegraphics[width= 0.8\linewidth,clip=]{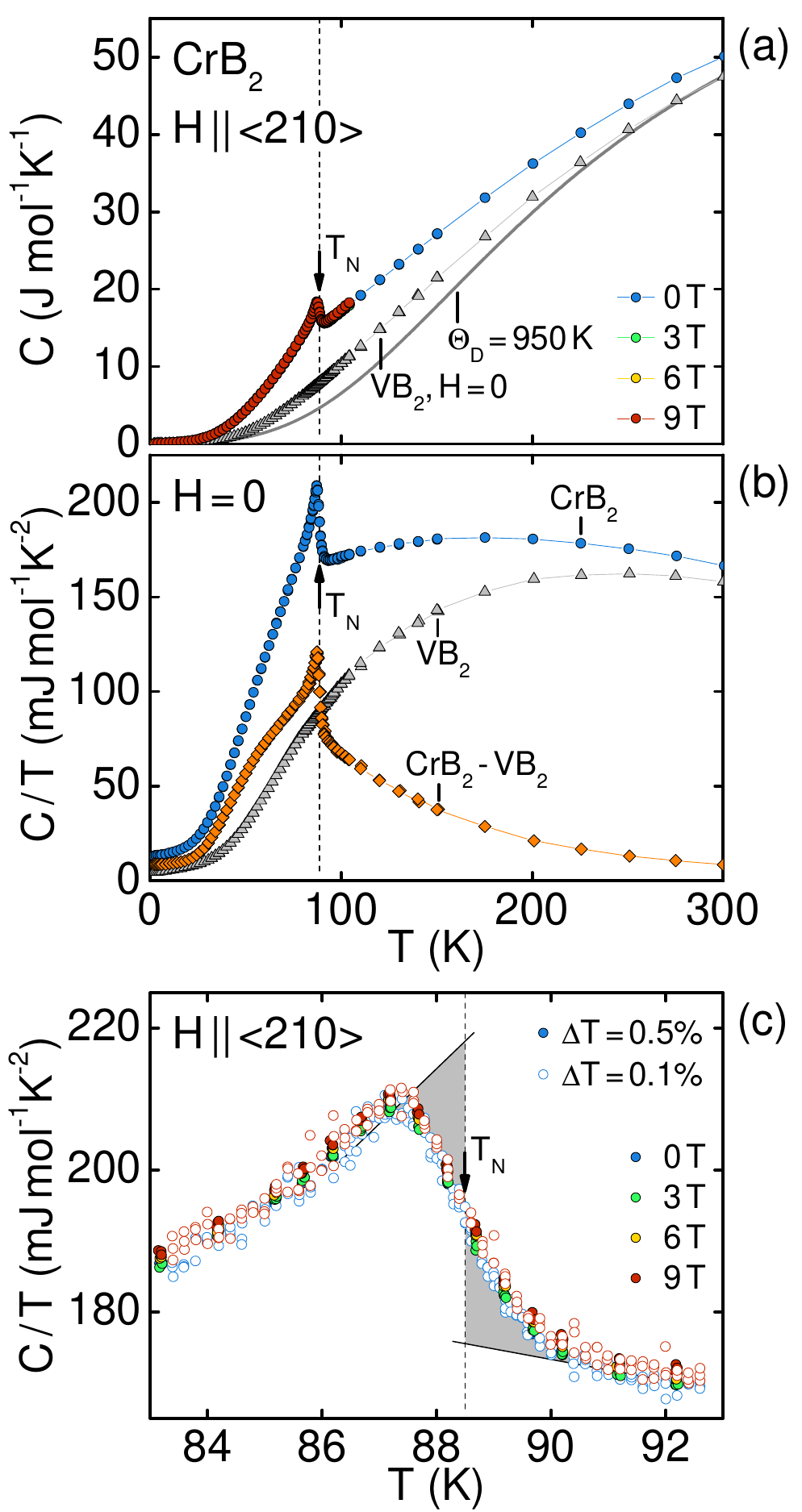}
\caption{(Color online) Specific heat of {\crb} as a function of temperature. (a)~A pronounced anomaly is present at the N$\acute{\rm{e}}$el transition. It hardly changes in fields up to 9\,T. The Dulong-Petit limit corresponds to $9R = 74.83$\,J\,mol$^{-1}$K$^{-1}$. (b)~Specific heat divided by temperature, $C/T$. Data are shown for {\crb} and VB$_{2}$, where the latter is used for an estimate of the lattice contribution in {\crb}. (c)~Specific heat divided by temperature around the N$\acute{\rm{e}}$el transition in more detail. An entropy-conserving construction yields the N$\acute{\rm{e}}$el temperature $T_{\rm{N}} = 88.5$\,K. Measurements with heat pulses of 0.1\% of the current temperature (open symbols) reproduce data measured with 0.5\% heat pulses.}
\label{figure06}
\end{figure}

\subsection{Specific Heat}
\label{specificheat}
 
Shown in Fig.~\ref{figure06} is the specific heat of our {\crb} sample. The limit of Dulong-Petit corresponds to $C_{\rm DP}=9R = 74.83\,{\rm J\,mol^{-1}K^{-1}}$. For low temperatures, where lattice contributions to the specific heat freeze out, we observe a Sommerfeld coefficient to the specific heat $\gamma_{0} = 13\,{\rm mJ\,mol^{-1}\,K^{-2}}$, consistent with the limited low-temperature data shown in the literature~\cite{Castaing1969}. This value is larger than for nonmagnetic diborides but typical for $d$-electron systems with moderate or strong electronic correlations~\cite{Vajeeston2001,Michioka2007}. Up to 9\,T, the highest field studied, we do not observe a field dependence of the low-temperature specific heat. This contrasts the pronounced suppression on similar field scales observed in a wide range of $d$- and $f$-electron systems also known as field-induced quenching of spin fluctuations.

With increasing temperature the high density of the data recorded in our study reveals a clear lambda anomaly at the N$\acute{\rm{e}}$el transition, characteristic of second-order mean-field behavior. As for low temperatures no changes are observed in fields up to 9\,T, where the high density of our data shows that the field dependence of $T_{\rm N}$ (if any) must be less than a tenth of a K. Measurements with two different sizes of heat pulses, 0.1\% and 0.5\%, are shown in Fig.~\ref{figure06}(c). Both lead to the same behavior underscoring that the observed detailed temperature dependence of the anomaly is not affected by the experimental method and intrinsic to the sample. Using a conventional entropy conserving construction (gray shaded areas) provides a N$\acute{\rm{e}}$el temperature $T_{\rm{N}} = 88.5$\,K in remarkable agreement with the resistivity and magnetization measurements. It is important to emphasize that the width of the transition of roughly $\pm1.5\,{\rm K}$ corresponds to $\sim\pm1.5\%$, which is actually quite small and perfectly consistent with the excellent sample quality seen in the other properties. 

Since the N$\acute{\rm{e}}$el temperature $T_{\rm N}\approx88.5\,{\rm K}$ is rather high, lattice contributions do not allow us to infer formation on the (spin wave) gap, seen in the resistivity, directly from the temperature dependence of the specific heat. In order to obtain at least a rough estimate of the lattice contributions to the specific heat of {\crb} we have measured the specific heat of VB$_2$, also shown in Fig.~\ref{figure06}(a). We justify this approach with the layered crystal structure and the very stable chemical bonding as an outstanding characteristic that is common to all C32 diborides. In contrast, a more detailed estimate of the lattice contribution to the specific heat in {\crb} inferred from first principles is well beyond the scope of our study. This may be illustrated by a simple estimate of lattice contributions to the specific heat based on a Debye model with a Debye temperature $\Theta_{\rm D}=950\,{\rm K}$ similar to that reported for other diborides, where the Debye temperature was chosen such that the slope $\text{d} C / \text{d}T$ approaches that of {\crb} and VB$_2$ at high temperatures. As evident from Fig.~\ref{figure06}(a) the clear difference with the specific heat of nonmagnetic VB$_2$ suggests for intermediate temperatures the presence of additional phonon contributions in comparison to a simple Debye behavior, that are most likely due to the reduced dimensionality of the hexagonal crystal structure.

Considering now the difference of the specific heat of {\crb} and VB$_2$, shown in Fig.~\ref{figure06}(b), several unusual features may be noted that are insensitive to the precise quantitative value of the specific heat of VB$_2$. First, the Sommerfeld contribution to the specific heat of VB$_2$ of $\gamma_{0} = 4\,{\rm mJ\,mol^{-1}\,K^{-2}}$ is comparatively large. Second, for $T>T_{\rm N}$ the specific heat of {\crb} includes substantial magnetic contributions pointing at the presence of very strong spin fluctuations. This observation is qualitatively consistent with the large effective fluctuating moment inferred from the Curie-Weiss dependence of the magnetization. Third, for $T<T_{\rm N}$ the magnetic contribution to the specific heat displays negative curvature around $50\,{\rm K}$; i.e., the temperature dependence is clearly more complex than the exponential dependence anticipated for a straight-forward formation of a spin wave gap. 

In fact, in a large number of systems the temperature dependence of the specific heat follows qualitatively the derivative of the resistivity. This may be expected theoretically, when the scattering seen in transport follows Fermi's golden rule, where the density of states dominates the specific heat. A comparison of $C(T)/T$ [Fig.~\ref{figure06}(b)] with $\text{d}\rho / \text{d}T$ [Fig.~\ref{figure03}(b)] underscores the positive curvature at intermediate temperatures, pointing at the presence of additional excitations that do not affect the resistivity. One possible origin of these excitations may be related to the effects of strong geometric frustration inferred from the paramagnetic susceptibility. Alternatively, the temperature dependence of the magnetic contribution to the specific heat in {\crb} appears to be reminiscent of features in the specific heat of UGe$_2$, where they have inspired speculations on an incipient charge density wave instability. The microscopic origin of such a putative coupled spin and charge density wave order, however, may be completely different for UGe$_2$ and {\crb}, notably rather one-dimensional aspects of the crystal structure in the former case as opposed to nesting-driven spin and charge density wave order as in pure Cr for the latter case. 

\begin{figure}
\includegraphics[width= 0.8\linewidth,clip=]{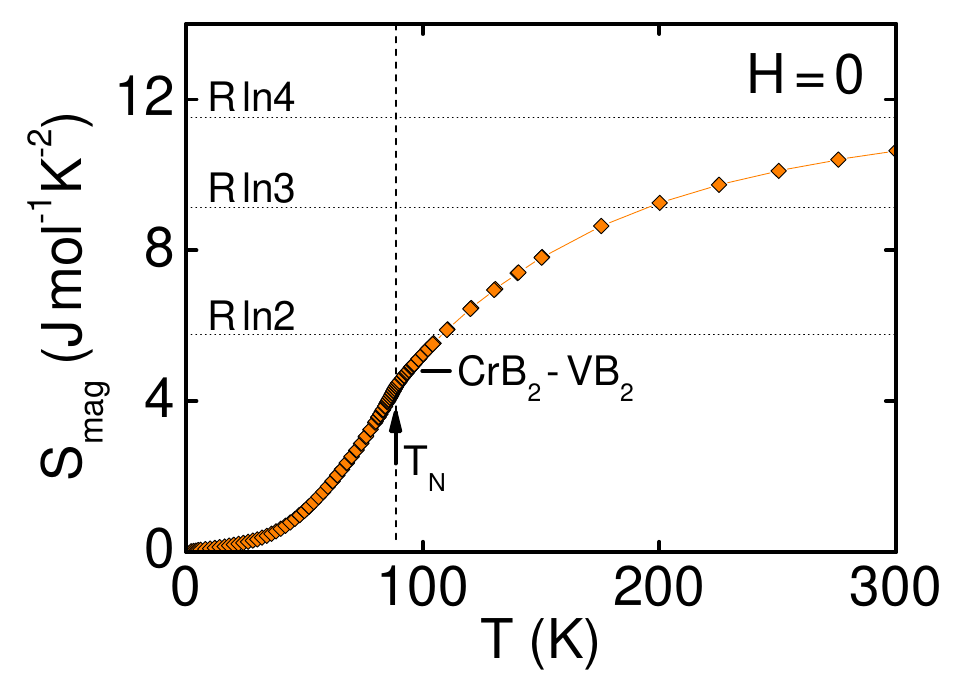}
\caption{(Color online) Magnetic contribution to the entropy of {\crb}. The lattice contribution was inferred from the entropy of VB$_2$ as described in the text.
}
\label{figure07}
\end{figure}

As a final aspect we present in Fig.~\ref{figure07} an estimate of the temperature dependence of magnetic contributions to the entropy of {\crb}. Here, we have used the specific heat of VB$_2$ as estimate of lattice contributions in {\crb} and integrated the remaining part, $\Delta C/T$, numerically. With increasing temperature the entropy appears to approach $R\ln 4$, which suggests a Cr$^{3+}$ state. Unfortunately, it is presently not possible to provide a microscopic underpinning for this conjecture, since the existing band structure calculations cannot capture the physics of fluctuating moments. A more careful account of the magnetic entropy released at high temperatures and its consistency with the fluctuating Curie-Weiss moment therefore has to await further experimental and theoretical exploration. 


\section{Discussion}
\label{further}

In this section we discuss the broader implications of our experimental findings. We comment at first on the lack of magnetic field dependence in all properties studied, which provides unambiguous evidence of weak itinerant antiferromagnetism. This is followed by a discussion of the consistency of the properties of {\crb} with an antiferromagnetically spin-polarized Fermi liquid ground state and the Kadowaki-Woods ratio in other systems. We then turn to the normal metallic state above $T_{\rm N}$, where we find broad consistency with a logarithmic temperature dependence of the specific heat suggesting two-dimensional antiferromagnetic spin fluctuations in a three-dimensional host. The discussion concludes with a short speculative note on the potential to observe superconductivity when {\crb} is forced to undergo an antiferromagnetic quantum phase transition, e.g., in high-pressure studies.

\subsection{Magnetic Field Dependence}

The effects of applied magnetic field, which to the best of our knowledge have not been addressed in any of the previous studies, are remarkably small up to 14\,T, the highest field studied. While the electrical resistivity near $T_{\rm N}$ shows a small quadratic increase of a few \% up to 14\,T, there is in particular no change of the anomaly and temperature of the magnetic transition. The same is observed in the magnetization and specific heat, providing consistent information.

Even though the magnetic energy associated with a field of 14\,T may still appear to be small as compared with $T_{\rm N}$, the Weiss temperature, and the spin fluctuation temperature, it is large as compared with the accuracy of a tenth of a K, at which our studies consistently reveal that there are no changes. This observation is even more striking when noticing that the evidence for strong fluctuations imply an inherent softness of the magnetic properties. Taken together the underlying energy scales, which drive the magnetic order, must be very large and typical of band structure effects. Namely, the field-induced Zeeman splitting of the conduction bands at 14\,T is still tiny; i.e., the magnetic order must be due to itinerant electrons. 

While the evidence for a gapped Fermi liquid ground state and the reduced ordered moment as compared with the fluctuating Curie-Weiss moment already hint at itinerant antiferromagnetism, we argue that the lack of field dependence provides unambiguous evidence for weak itinerant antiferromagnetism in {\crb}.

\subsection{Kadowaki-Woods Ratio}

An empirical classification that considers the consistency of a wide range of materials exhibiting strong electronic correlations with Fermi liquid theory is the so-called Kadowaki-Woods ratio, i.e., the ratio of the coefficient of the quadratic temperature dependence of the resistivity, $A$, to the square of the coefficient of the linear temperature dependence of the specific heat, $\gamma^2$~\cite{Kadowaki:SSC1986}. While the former may be interpreted as a probe of the cross section in quasiparticle-quasiparticle scattering, the latter represents the effective quasiparticle mass (the former probes the imaginary part of the self-energy, while the latter probes the real part). In turn, the Kadowaki-Woods ratio is only meaningful when the self-energy is essentially momentum independent.

\begin{figure}
\includegraphics[width= 0.75\linewidth,clip=]{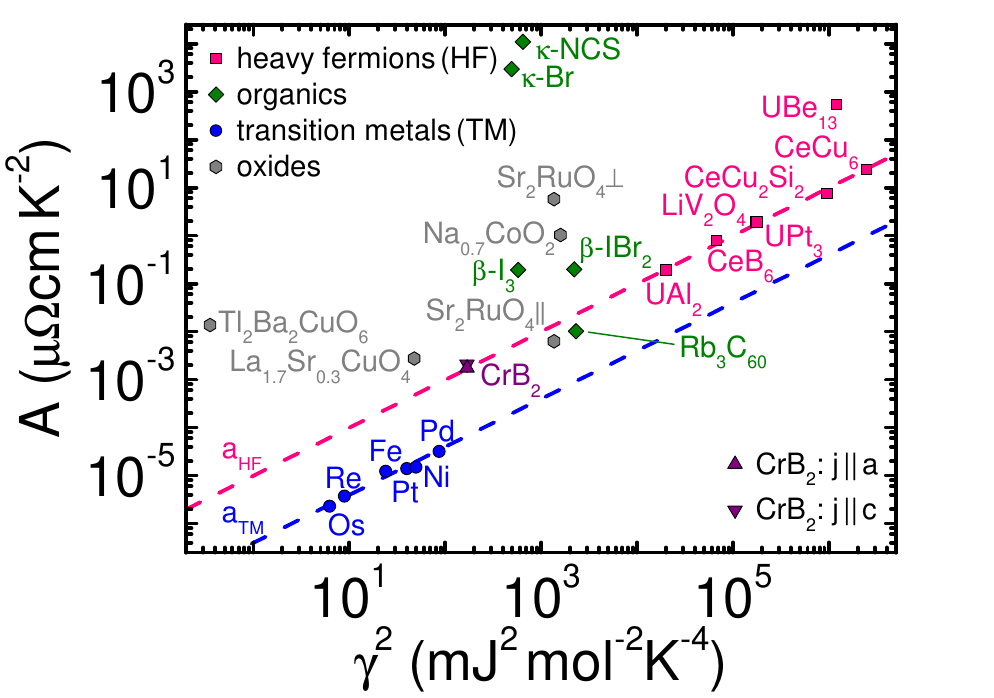}
\caption{(Color online) Kadowaki-Woods plots of selected $d$- and $f$-electron compounds as presented in Ref.~\onlinecite{Jacko2009}. The Kadowaki-Woods ratio of {\crb} is close to the value $a_{\rm HF} = 10\,{\rm \mu\Omega\,cm\,mol^{2}K^{2}J^{-2}}$ observed in heavy-fermion compounds (upper dashed line).
}
\label{figure08}
\end{figure}

It is interesting to note that the quantitative values of the coefficients $A$ and $\gamma$ that we infer from the resistivity and the specific heat are typical of other $d$-electron systems~\cite{Pfleiderer2002,Boeuf2006,Pfleiderer:JMMM2001,Bauer:PRB2010}. The values of the coefficient $A$ were determined in the presence of antiferromagnetic order taking into account the scattering by impurities and the formation of a spin wave gap. The corresponding contribution in the specific heat is given by $\gamma_{0} = 13\,{\rm mJ\,mol^{-1}\,K^{-2}}$. In view of the finite transition temperature and the putative evidence for a gap, the assumption of momentum independence appears justified. As shown in Fig.~\ref{figure08}, the values of $A$ and $\gamma$ then correspond to a Kadowaki-Woods ratio typical of $f$-electron heavy fermion systems, while it differs clearly from transition-metal elements. The values of {\crb} are thereby located in the regime of the cuprate superconductor La$_{1.7}$Sr$_{0.3}$CuO$_4$ and the ruthenate superconductor Sr$_2$RuO$_4$. 

\begin{figure}
\includegraphics[width= 0.75\linewidth,clip=]{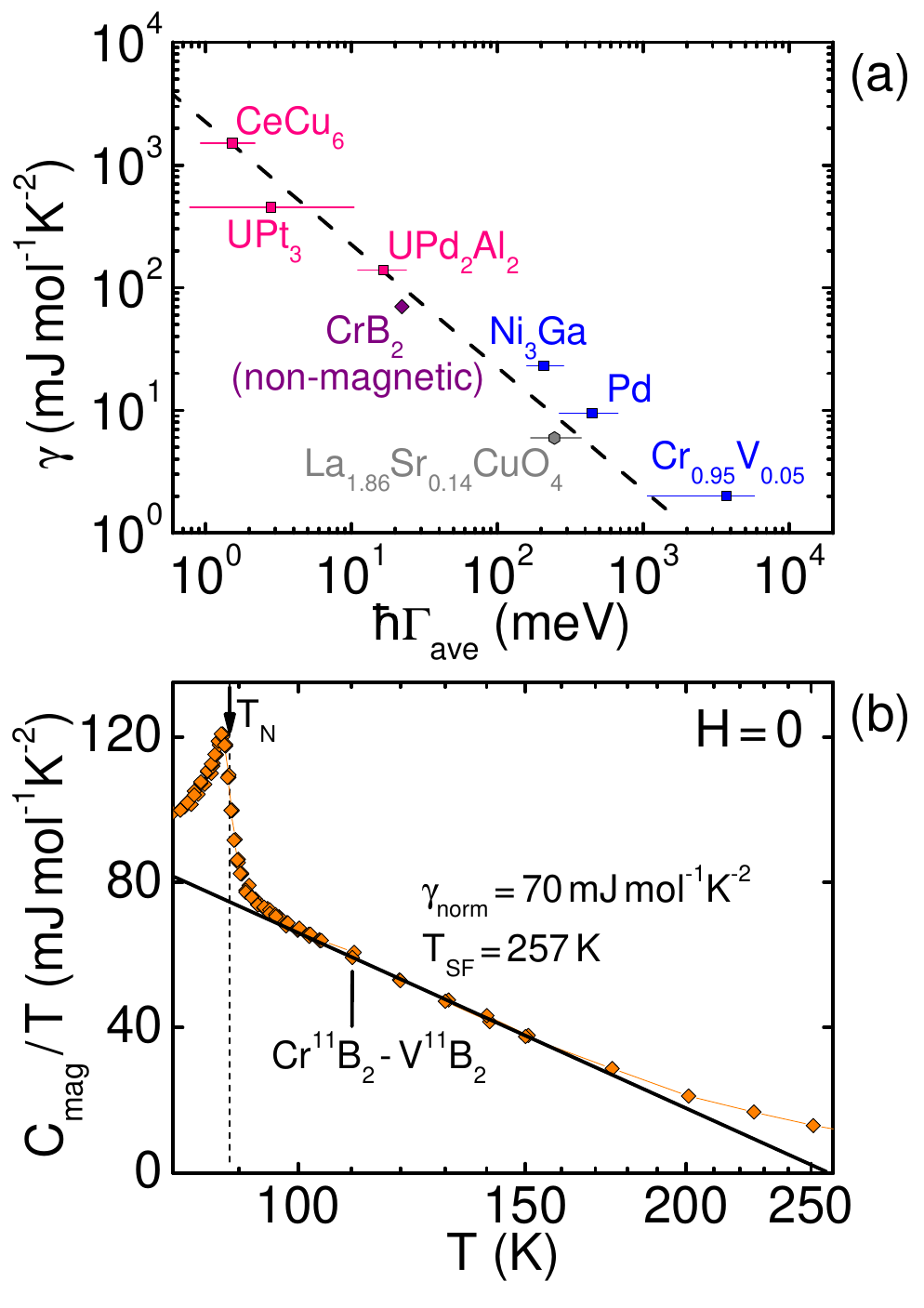}
\caption{(Color online) Estimate of the average fluctuation rate for nonmagnetic {\crb}. (a)~Correlation between the Sommerfeld coefficient, $\gamma$, and the average fluctuation rate, $\Gamma_{\rm ave}$, for various compounds. The figure has been reproduced from Ref.~\onlinecite{Hayden2000}. An estimate for nonmagnetic {\crb} is inferred from the specific heat in the following panel. (b)~Magnetic contribution to the specific heat of {\crb} at high temperatures. A fit corresponding to $C_{\rm mag}/T=\gamma_{\rm norm}\ln(T_{\rm SF}/T)$ (solid line) yields an estimate of the Sommerfeld coefficient expected for nonmagnetic {\crb}, $\gamma_{\rm norm} = 70\,{\rm mJ\,mol^{-1}\,K^{-2}}$, and a spin fluctuation temperature, $T_{\rm SF} = 257$\,K.
}
\label{figure09}
\end{figure}

\subsection{Spin Fluctuations}

Among the bulk properties of {\crb} reported in this paper the large fluctuating Curie-Weiss moment and the large magnetic contribution to the specific heat above $T_{\rm N}$ provide consistent evidence for an abundance of strong spin fluctuations in the normal state. In the framework of the self-consistently renormalized Ginzburg-Landau theory of itinerant-electron magnets the dynamical susceptibility associated with such a relaxation frequency spectrum is parametrized in terms of a single-pole function at the magnetic ordering wave vector. Material-specific parameters describe the damping and spin wave stiffness. Assuming this ansatz for the dynamical susceptibility, notably unique parameters of the damping and spin wave stiffness across the Brillouin zone, the normal state contribution to the specific heat scales with an average fluctuation rate $\Gamma_{\rm ave}$. Empirically this has been demonstrated by Hayden \textit{et al.}~\cite{Hayden2000} for a wide range materials comprising ferromagnetic and antiferromagnetic compounds as reproduced in Fig.~\ref{figure09}(a). The average fluctuation rate may thereby be expressed in terms of a material specific spin fluctuation temperature $\hbar\Gamma_{\rm ave}=k_{\rm B}T_{\rm SF}$. 

In the spirit of the framework of the self-consistently renormalized spin fluctuation theory the specific heat of {\crb} for $T > T_{\rm N}$ provides an estimate of the characteristic temperature of the fluctuations in the normal state of {\crb}, $T_{\rm SF}$, and of the corresponding prefactor $\gamma_{\rm norm}$. Assuming two-dimensional fluctuations in a three-dimensional system the normal-state specific heat is thereby expected to vary as $C_{\rm mag}/T=\gamma_{\rm norm}\ln(T_{\rm SF}/T)$. We justify the consideration of two-dimensional fluctuations by the layered crystal structure of {\crb}, the weak easy-plane magnetic anisotropy, and the anisotropic resistivity. In addition, the expression for three-dimensional antiferromagnetic fluctuations, $C/T=\gamma-D\sqrt{T}$, empirically does not account for the data as well.

Shown in Fig.~\ref{figure09}(b) is the magnetic contribution to the specific heat over temperature on a logarithmic temperature scale, where we find reasonable agreement with the theoretical expression, i.e., a logarithmic temperature dependence. The deviation at the highest temperatures may be due to the inaccuracy of the subtraction of the lattice contribution. We find for the normal state a Sommerfeld coefficient of $\gamma_{\rm norm} \approx 70\,{\rm mJ\,mol^{-1}\,K^{-2}}$ and a spin fluctuation temperature $T_{\rm SF} \approx 257\,{\rm K}$, which corresponds to $\hbar\Gamma_{\rm ave} = k_{\rm B}T_{\rm SF} \approx 22\,{\rm meV}$. This ratio of $\gamma_{\rm norm}$ and $\Gamma_{\rm ave}$ in the normal state of {\crb} agrees remarkably well with the empirical observation by Hayden \textit{et al.} shown in Fig.~\ref{figure09}(a), hence suggesting that the assumptions made above are justified.

\subsection{Further Aspects}

The possible occurrence of spin fluctuation mediated superconductivity has been discussed in a wide range of materials; see, e.g., Ref.~\onlinecite{Monthoux2007}. The superconducting instability requires a number of factors to be satisfied, namely, (i)~the spin fluctuation spectra should be fairly well focused in energy and momentum, (ii)~the dominant momentum contributions should match sections of the Fermi surface as to promote superconductive pairing, and (iii)~transverse and longitudinal components should be pair forming. The superconducting transition temperature may then be expected to scale with the width of the relaxation frequency spectrum as measured by a characteristic spin fluctuation temperature $T_{\rm SF}$. 

Following this line of thought it has long been pointed out that the heavy-fermion and cuprate superconductors may be empirically related in a plot of the superconducting transition temperature $T_{\rm c}$ against the spin fluctuation temperature $T_{\rm SF}$~\cite{Moriya:RPP2003,Pfleiderer2009}. Assuming now that all microscopic factors promoting magnetically mediated superconductivity may indeed be satisfied for \textit{nonmagnetic} {\crb}, this empirical relationship suggests a superconducting transition temperature as high as $T_{\rm c}\sim10\,{\rm K}$ for {\crb}. 

It is important to stress that this consideration presumes that the antiferromagnetic order in {\crb} has been suppressed, e.g., through the application of hydrostatic or uniaxial pressure and that the normal-state properties at the zero temperature border of antiferromagnetism may be inferred from the normal state at ambient pressure as outlined above. In other words, such a prediction ignores, for instance, the possibility of the strong geometric frustration in {\crb} pointed out in Sec.~\ref{magnetization}.

\section{Conclusions}
\label{conclusions}

In conclusion, we have prepared high-quality single-crystal $^{11}$B-enriched {\crb} by a solid-state reaction of Cr and B powder and optical float zoning. The excellent quality of our samples is evident from (i)~the shiny, faceted appearance of the float-zoned ingot, (ii)~Laue x-ray and neutron diffraction, (iii)~the highest residual resistivity ratios reported in the literature to date, (iv)~the absence of a Curie tail in the magnetization, as well as a very well developed kink at the magnetic transition and quantum oscillations~\cite{Brasse:PRB2013}, and finally (v)~a well-defined specific-heat anomaly at the antiferromagnetic transition. 

The electrical resistivity, $\rho_{\rm xx}$, Hall effect, $\rho_{\rm xy}$, and specific heat, $C$, as well as the absence of a divergence of the magnetization for $T\to0$, are characteristic of an exchange-enhanced Fermi liquid ground state, which develops a slightly anisotropic spin gap below $T_{\rm N}$. The absence of a divergence in the magnetization for $T\to0$, reported in previous studies of {\crb}, rules out an incipient marginal Fermi liquid state as observed for weakly ferromagnetic materials. Comparison of $C$ with $d\rho_{\rm xx}/dT$ reveals a pronounced second-order mean-field transition at $T_{\rm N}$, and an abundance of antiferromagnetic spin fluctuations as well as unusual excitations below $T_{\rm N}$. The anisotropy of the magnetization and $\rho_{\rm xx}$ are characteristic of a weak easy-plane anisotropy of the magnetic properties as well as the spin fluctuations. 

We finally note that the ratio of the Curie-Weiss to the N$\acute{\rm{e}}$el temperatures, $f=-\Theta_{\rm CW}/T_{\rm N}\approx 8.5$, taken from the magnetization indicates strong geometric frustration. However, despite the evidence for strong spin fluctuations and geometric frustration, which imply an inherent softness of the magnetic properties to applied magnetic fields, all physical properties are remarkably invariant under applied magnetic fields as compared to the highest field studied of 14\,T. This property is clearly characteristic of itinerant antiferromagnetism and perhaps most remarkable in its own right. In contrast to earlier suggestions of local-moment magnetism our study hence identifies {\crb} as a weak itinerant antiferromagnet with strong geometric frustration, providing the perhaps closest antiferromagnetic analog to the weak itinerant ferromagnets reported so far.

\acknowledgements
We wish to thank G. Behr, B. B\"uchner, M. Brasse, L. Chioncel, A. Erb, C. Franz, D. Grundler, M. Halder, W. L\"oser, W. Kreuzpaintner, S. Mayr, C. Morkel, M. Schulze, A. Senyshyn, M. Vojta, M. Wagner, and M. Wilde, as well as the research technology department of the IFW Dresden for fruitful discussions and technical support. Part of this work is based upon experiments performed at the HEiDI instrument operated by RWTH Aachen/FZ J\"ulich (J\"ulich Aachen Research Alliance, JARA) at the Heinz Maier-Leibnitz Zentrum (MLZ), Garching, Germany. A.B. and A.R. acknowledge support through the TUM Graduate School. S.W. acknowledges support by the Deutsche Forschungsgemeinschaft (DFG) under the Emmy-Noether program in Project No.\ WU595/3-1. Financial support through DFG TRR80 is gratefully acknowledged.


\begin{thebibliography}{77}
\bibitem[{neu()}]{neutron-absorption}
\bibinfo{note}{$^{10}$B displays a very high neutron absorption cross section
  $\sigma_{\rm A} = 3835\,{\rm b}$ for thermal neutrons as compared with that
  of $^{11}$B and $^{\rm{nat}}$Cr which have $\sigma_{\rm A}=0.0055\,{\rm b}$
  and $3.05\,{\rm b}$, respectively. Natural boron displays an isotope
  composition of $19.9\% ^{10}$B and $80.1\% ^{11}$B. The resulting absorption
  coefficient $\Sigma^{\rm{abs}}_{\rm{nat}}= 65.2$\,cm$^{-1}$ of CrB$_{2}$
  prepared with natural boron therefore implies a transmission of $1.5
  \cdot10^{-3}$ and $4.5\cdot10^{-14}$ of the initial neutron intensity through
  a CrB$_{2}$ sample of 1\,mm thickness and a cylindric sample of 6\,mm
  diameter, respectively. For the 99\,\% isotopically enriched $^{11}$B used in
  our study this absorption coefficient reduces to
  $\Sigma^{\rm{abs}}_{\rm{B11}}= 3.40$\,cm$^{-1}$. As a result the transmission
  ratio increases to 0.71 and 0.20 for the 1\,mm plate and 6\,mm cylinder,
  respectively.}
\end{thebibliography}

\end{document}